\documentclass{epl}
\usepackage{amsmath,amssymb}
\title{Lateral phase separation in mixtures of lipids
and cholesterol systems} \shorttitle{lipid-cholesterol}
\author{Shigeyuki Komura\inst{1}
\thanks{E-mail: \email{komura@comp.metro-u.ac.jp}},
        Hisashi Shirotori\inst{1},
        Peter D. Olmsted\inst{2},
        \and
        David Andelman\inst{3}}
\institute{
  \inst{1} Department of Chemistry, Faculty of Science,
           Tokyo Metropolitan University,
           Tokyo 192-0397, Japan \\
  \inst{2} School of Physics and Astronomy,
           University of Leeds, Leeds LS2 9JT, UK \\
  \inst{3} School of Physics and Astronomy,
           Raymond and Beverly Sackler Faculty of Exact Sciences,
           Tel Aviv University, Ramat Aviv 69978, Tel Aviv, Israel
}
\pacs{87.16.Dg}{Membranes, bilayers, and vesicles}
\pacs{64.70.Ja}{Liquid-liquid transitions}
\pacs{64.75.+g}{Solubility, segregation, and mixing; phase separation}

\begin{document}

\maketitle
\begin{abstract}
  In an effort to understand ``rafts'' in biological membranes, we
  propose phenomenological models for saturated and unsaturated
  lipid mixtures, and lipid-cholesterol mixtures.  We consider
  simple couplings between the local composition and internal
  membrane structure, and their
  influence on transitions between liquid and ``gel'' membrane phases.
  Assuming that the gel transition temperature of the saturated lipid
  is shifted by
  the presence of the unsaturated lipid, and that  cholesterol
  acts as an external field on the chain melting transition,
  a variety of phase diagrams are obtained.
  The phase
  diagrams for binary mixtures of
  saturated/unsaturated lipids and lipid/cholesterol are
  in semi-quantitative agreement with the experiments.
  Our results also apply to regions in the ternary phase diagram of
  lipid/lipid/cholesterol systems.
\end{abstract}

\section{Introduction}

Biological membranes typically contain various components such as
lipid mixtures, sterols, and proteins that are indispensable to cell
functions~\cite{LS}.  Rather than being uniformly distributed in the
membrane, there is growing evidence that some intra-membrane
components are incorporated in domains arising from lateral lipid
segregation in membranes. This phenomena has attracted great interest
in the context of ``rafts''~\cite{SI}, \textit{i.e.}, liquid domains
rich in cholesterol, saturated lipids (typically sphingomyelin
lipids), and in some cases particular proteins~\cite{BL}.  Moreover,
cholesterol-rich domains have been directly observed in model membranes
composed of lipid mixtures and cholesterol, using advanced
fluorescence microscopy~\cite{KSWF,DBVTLJG,SGL,FB,VK,BHW,GGKJGKJ}.

Prior to the notion of ``rafts'' in biological membranes, the role
of cholesterol was investigated in binary lipid-cholesterol
membranes~\cite{BM,IKMWZ,NMIZM}, where phase separation was
observed using NMR and calorimetry~\cite{VD,ST}. Lipid membranes
undergo a freezing or ``gel-like'' transition, in which the
hydrocarbon tails order. Addition of cholesterol has several
effects. It suppresses the ``gel'' transition below physiologically
relevant temperatures, and can lead to coexistence of
two liquid phases with  very different orientational order. It is now
believed that model membranes containing two phospholipids
(saturated and unsaturated) and
cholesterol~\cite{KSWF,DBVTLJG,SGL,FB,VK,BHW,GGKJGKJ} exhibit  ``rafts''
which are \textit{liquid-ordered} (\textit{L$_o$})
domains, coexisting with a surrounding  background in a
\textit{liquid-disordered} (\textit{L$_d$}) state~\cite{BL}.

From a physical viewpoint, a strategy for understanding the basic
structure of ``rafts'' in biological membranes is as
follows~\cite{LD}. First, it is necessary to have a simple model
for binary mixtures of saturated and unsaturated lipids. Second, a
minimal model describing binary lipid-cholesterol systems is
required in order to understand the effects of cholesterol on
membrane phase behavior.  Finally, these two viewpoints could be
combined to fully investigate three-component systems. Here we
focus on the first two steps, and propose simple phenomenological
models for lipid-lipid and lipid-cholesterol binary mixtures. Such
an approach is quite useful since we can predict, at least
qualitatively, the complex phase behavior [shown, \textit{e.g.},
in fig.~\ref{schematic}(a)] of the ternary lipid-lipid-cholesterol
system from the three binary sub-systems.

It is generally believed that the gel phase transition is driven by
the freezing of lateral motion as well as conformational ordering of
lipids. However, because these two degrees of freedom are strongly
coupled, the gel phase transition occurs at a single temperature for
pure lipid systems. One of our major assumptions is that the membrane
state, even for lipid mixtures, can be described by one internal
degree of freedom, coupled to the lateral phase separation. Our main
result is phase diagrams for two component systems, which reproduce
experimental ones, without specifying the detailed microscopic state
of the lipids. Moreover, we shed light on recent phase diagrams
obtained for ternary mixture \cite{VK,BHW}.

\section{Model for saturated-unsaturated lipids systems}

\begin{figure}[t]
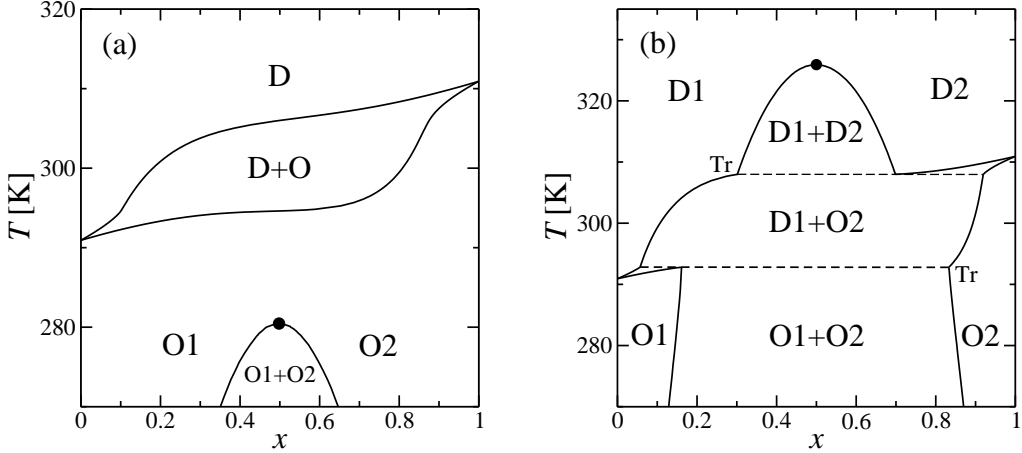

\twoimages[scale=0.35]{fig1a.eps}{fig1b.eps} \caption{Calculated
phase diagrams of binary mixtures of saturated and unsaturated
lipids as a function of mole fraction of the saturated lipid $x$
and temperature $T$ for (a) $J= 4.0 \, k_{\rm B} T^{\ast}_{\rm
s}$, and (b) $J= 5.0 \, k_{\rm B} T^{\ast}_{\rm s}$. The other
parameters are $a_2' = 174 \, k_{\rm B}, a_3 = -307 \, k_{\rm B}
T^{\ast}_{\rm s}, a_4 = 613 \, k_{\rm B} T^{\ast}_{\rm s}$,
$T^{\ast}_{\rm u} = 240$K, $T^{\ast}_{\rm s} = 260$K as estimated
in \cite{GL}.
The gel transition temperature is $T_{\rm g}=291$K and $311$K for
$x=0$ and $1$, respectively. The ordered (large $\psi$) and
disordered ($\psi=0$) phases are respectively denoted by O and D.
The critical point is indicated by a filled circle, and the triple
point by Tr. The critical points are located at (a) $x_{\rm c}=
0.498$, $T_{\rm c}=280$K, and (b) $x_{\rm c}= 0.5$, $T_{\rm
c}=326$K, respectively. } \label{lipidlipid}
\end{figure}

If different lipids exhibit only a small difference in their gel
transition temperature, their phase diagram in terms of
temperature-composition parameters will include a ``cigar-like"
shape of liquid-gel coexistence [as in fig.~\ref{lipidlipid}(a)].
As the gel transition temperature difference between the two
lipids is increased, a gel-gel coexistence region appears below
the liquid-gel coexistence. At even larger difference, the phase
diagram becomes more complex as the two coexistence regions
partially overlap~\cite{Sackmann}.

We consider a single bilayer membrane composed of $x$ mole
fraction of saturated lipid and $(1-x)$ of unsaturated lipid. The
two lipids are taken to have different gel transition temperatures
originating from different chain length, degree of saturation, or
hydrophilic head group. We assume the same area per molecule for
both species, and ignore lipid exchange with the surrounding
solvent.  The total free energy $f^{\ell\ell}=
f_1^{\ell\ell}+f_2^{\ell\ell}$ comprises (i) the free energy
$f^{\rm \ell\ell}_1$ of mixing and binary interactions, and (ii)
the ``stretching'' energy $f^{\rm \ell\ell}_2$, which controls
changes in bilayer thickness.

The free energy per site $f^{\rm \ell\ell}_1$ is the sum of the
entropy of mixing and enthalpy. It can be written within a
Bragg-Williams (mean-field) theory as
\begin{equation}
f^{\rm \ell\ell}_1(x) = k_{\rm B}T
\left[ x \log x + (1-x) \log (1-x) \right]
- \tfrac{1}{2} J x^2,
\label{eq:bragg}
\end{equation}
where $k_{\rm B}$ is the Boltzmann constant, $T$ is the
temperature, $J>0$ is an attractive interaction parameter that
enhances demixing. Linear terms in $x$ can be disregarded because
they merely shift the origin of the chemical potential.

To describe the gel transition, which involves chain ordering and
stiffening, we introduce a rescaled membrane thickness $\psi
\equiv (\delta-\delta_0)/\delta_0$ as an order parameter, where
$\delta$ is the actual membrane thickness and $\delta_0$ is the
constant membrane thickness in the disordered phase corresponding
to the liquid phase~\cite{GL}.
Note that $\psi$ summarizes changes in various degrees of freedom,
including the conformations of the hydrocarbon chains and
their inter-chain correlations. Since the gel transition is first
order, an appropriate phenomenological Landau expansion of the
``stretching'' free energy per site in powers of $\psi$
is~\cite{GL}
\begin{equation}
f^{\rm \ell\ell}_2(x,\psi) = \tfrac{1}{2} a_2'[T-T^{\ast}(x)] \psi^2 +
\tfrac{1}{3} a_3 \psi^3 + \tfrac{1}{4} a_4 \psi^4,
\label{eq:landau1}
\end{equation}
where $a_2'> 0$, $a_3 < 0$, and $a_4 > 0$. For a single component
membrane (\textit{i.e.}, $x=0$ or $1$), $T^{\ast}$ is a reference 
temperature,
and the first-order gel transition temperature is given by
$T_{\rm g} = T^{\ast} + 2 a_3^2/(9 a_2' a_4)$. For a binary mixture,
the transition temperature depends on the composition $x$; for
convenience we describe the reference temperature $T^{\ast}(x)$ as
a linear interpolation between the two pure limits:
\begin{equation}
T^{\ast}(x) = x T^{\ast}_{\rm s} + (1-x)T^{\ast}_{\rm u},
\label{eq:interpolate}
\end{equation}
where $T^{\ast}_{\rm s}$ and $T^{\ast}_{\rm u}$ are the reference
temperatures of the pure saturated and unsaturated lipids,
respectively. With all else the same (head group size,
interactions, and chain length, \textit{etc}.), we expect
$T^{\ast}_{\rm s} > T^{\ast}_{\rm u}$, because unsaturated lipids
break up the crystallizing tendencies. Note that
eq.~(\ref{eq:interpolate}) leads to a coupling term $x\psi^2$
\cite{priest}. We have neglected a lower order bilinear term $x
\psi$, which induces a small temperature and composition
dependence in $\delta_0$.  Another possible coupling term $x^2
\psi$ simply renormalizes the interaction parameter $J$.

Combining eqs.~(\ref{eq:bragg}-\ref{eq:interpolate}), we obtain the
total free energy $f^{\rm \ell\ell}$. After minimizing with respect to
$\psi$, the two-phase region is obtained by the Maxwell construction.
In fig.~\ref{lipidlipid}, we show two typical phase diagrams of binary
lipid mixtures.  For small $J$ (fig.~\ref{lipidlipid}(a)) which
corresponds experimentally to mixtures with weak segregation tendency,
a cigar-like coexistence region is obtained between a disordered (D)
phase where $\psi=0$ (liquid) and an ordered (O) phase where $\psi>0$
(gel).  At temperatures below the cigar-shape region, there is another
coexistence region between two ordered phases, O1+O2, with
$\psi_1\lesssim\psi_2$. This type of phase diagram was experimentally
observed for DEPC-DPPC binary lipid systems~\cite{WM}.  When $J$ is
increased, the O1+O2 coexistence region extends into and beyond the
D+O coexistence region (fig.~\ref{lipidlipid}(b)). In this more
complex phase diagram, there are two triple points (denoted as Tr) at
which the two disordered phases and ordered phase (D1+D2+O2), or the
disordered phase and two ordered phases (D1+O1+O2) coexist. Many
features of this phase diagram can be seen for DEPC-DPPE lipid
mixtures which have a stronger segregation tendency because both the
head and the tail moieties are different~\cite{WM}.

\section{Model for lipid-cholesterol systems}

\begin{figure}[t]
\oneimage[scale=0.6]{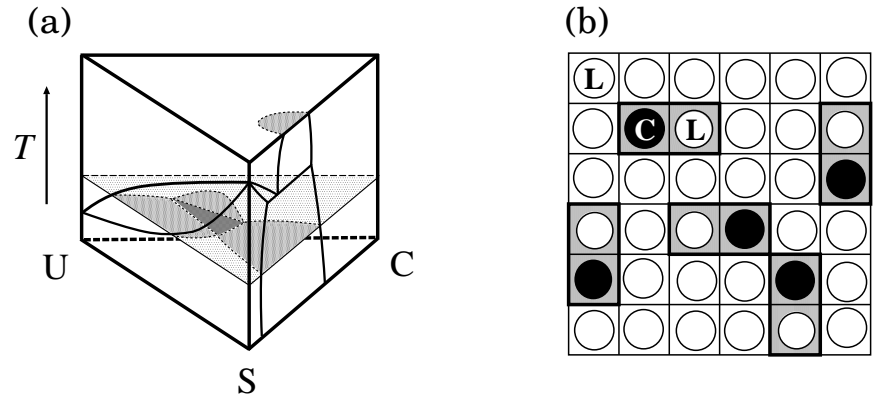}
\caption{ (a) Schematic phase prism
  of a ternary system consisting of saturated lipid (S), unsaturated
  lipid (U), and cholesterol (C). The gray and black regions on the
  constant temperature triangle plane are two-phase and three-phase
  coexisting regions, respectively. (b) Lattice model for a
  lipid-cholesterol mixture. Each cholesterol molecule ($\bullet$)
  forms a dimer with a neighboring lipid molecule ($\circ$).}
\label{schematic}
\end{figure}

Next we discuss the role of sterols, such as cholesterol and
lanosterol, on the phase behavior. On one hand, a small amount of
cholesterol destabilizes the gel phase in favor of a
\textit{liquid-disordered} (\textit{L$_d$}) phase~\cite{VD,ST}.
Substantial cholesterol, on the other hand, stabilizes a
\textit{liquid-ordered} (\textit{L$_o$}) phase in which the lipid
hydrocarbon tails are extended, but maintain high lateral mobility.
This reflects the dual molecular mechanism of a cholesterol molecule:
it can act as (i) an ``impurity'' and weakens the inter-lipid
interactions for ordering or (ii) a ``chain rigidifier'' and induces
conformational order in neighboring lipid chains~\cite{NMIZM}.
Moreover, recent experiments using atomic force microscopy showed that
the cholesterol-rich domains are thicker than the cholesterol-poor
regions~\cite{YFBJ,LSEH}, supporting our assumption that the local
membrane thickness can be taken as the order parameter.

\begin{figure}[t]
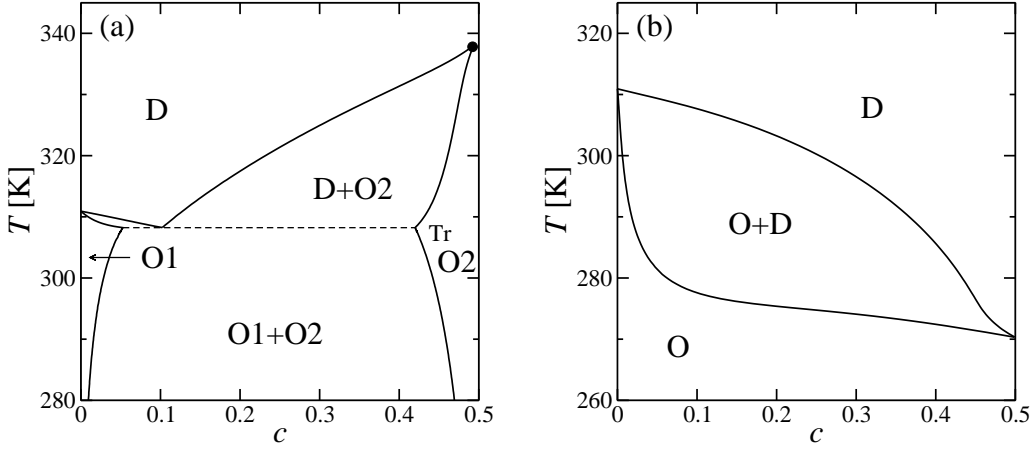

\twoimages[scale=0.35]{fig3a.eps}{fig3b.eps}
\caption{Phase diagrams of binary lipid-sterol mixtures as a
  function of mole fraction $c$ of sterol and temperature $T$ for (a)
  cholesterol, $\Gamma_1 = 11 \, k_{\rm B}T^{\ast}$, $\Gamma_2 = 47 \,
  k_{\rm B}T^{\ast}$; and (b) lanosterol, $\Gamma_1 = 28 \, k_{\rm
    B}T^{\ast}, \Gamma_2 = 20 \, k_{\rm B}T^{\ast}$. The other
  parameters are the same as in fig.~\ref{lipidlipid}, except
  $T^{\ast} = 260$K.  The gel transition temperature at $c=0$ is
  $T_{\rm g}=311$K.  The ordered (large $\psi$) and disordered (
  non-zero small $\psi$) phases are respectively denoted by O and D.  
  The critical point indicated by a filled circle in (a) is located 
  at $c_{\rm c}= 0.492$, $T_{\rm c}=338$K.}
\label{lipidchol}
\end{figure}

Based on these observations, we propose a model for
lipid-cholesterol binary mixtures. Consider a membrane with
cholesterol mole fraction $c$ and lipid mole fraction $(1-c)$.
There are three contributions to the total free energy $f^{\rm \ell
c}=f^{\rm \ell c}_1 + f^{\rm \ell c}_2 + f^{\rm \ell c}_3$: (i)
the entropy of mixing $f^{\rm \ell c}_1(c)$ between lipid and
cholesterol; (ii) the ``stretching'' energy $f^{\rm \ell
c}_2(\psi)$ of lipid molecules; and (iii) direct coupling terms
between lipid and cholesterol, $f^{\rm \ell c}_3(c,\psi)$, which
takes into account the effect of cholesterol on the bilayer
thickness.

We note that the coexistence region in the lipid-cholesterol
system is typically found only for $c \le 0.5$~\cite{VK,VD,ST}.
For cholesterol concentration above roughly 0.5, the entire
bilayer becomes unstable. To account for such observations we
consider ``condensed complexes'', following \cite{RAM}, and assume
that each cholesterol molecule in the membrane irreversibly
dimerizes with a single lipid molecule. Since all cholesterol
molecules form dimers, as shown in fig.~\ref{schematic}(b), the
entropy of mixing is taken between lipid-cholesterol dimers and
the remaining monomeric lipids, and is given by a Flory-Huggins
form:
\begin{equation}
f^{\rm \ell c}_1(c) = k_{\rm B}T \left[
c \log 2c + (1-2c) \log (1-2c) \right].
\label{eq:flory}
\end{equation}
This free energy is valid only for $0 \le c \le 0.5$, since each
cholesterol molecule is paired with a neighboring lipid. The first
term is the entropy of dimers having an area fraction of
$2c$, while the second term accounts for the entropy of lipid
monomers of area fraction $(1-2c)$. Although extensions of
the model to include excess free cholesterol and/or formation
of trimers, \textit{etc.} is
possible~\cite{SM}, the present situation is enough to describe
the qualitative phase behavior.

The Landau free energy describing the structural phase transition
of the membrane is given by $f^{\rm \ell c}_2(\psi)$ and is the same as
$f^{\rm \ell \ell}_2$ of eq.~(\ref{eq:landau1}):
\begin{equation}
f^{\rm \ell c}_2(\psi) = \tfrac{1}{2} a_2'(T-T^{\ast}) \psi^2 +
\tfrac{1}{3} a_3 \psi^3 + \tfrac{1}{4} a_4 \psi^4,
\label{eq:landau2}
\end{equation}
where the order parameter $\psi$ is again the relative bilayer
thickness, and $T^{\ast}$ is the reference temperature of the pure
lipid system.

The simplest free energy to account for the previously mentioned dual
effects of cholesterol can be expressed by the following
phenomenological coupling terms:
\begin{equation}
f^{\rm \ell c}_3(c,\psi) = \tfrac{1}{2} \Gamma_1 c \psi -
\tfrac{1}{2} \Gamma_2 c^2 \psi,
\label{eq:coupling}
\end{equation}
where $\Gamma_1 >0$ and $\Gamma_2 >0$ are the coupling constants.
The first term expresses the fact that a small amount of
cholesterol ($c>0$) acts as an ``impurity''. It interferes with the
crystalline ordering, and enhances disordered chains (smaller
$\psi$). This term is necessary because, as shown in the
experimental phase diagram~\cite{VD,ST}, the gel-\textit{L}$_d$
coexistence temperature decreases upon adding cholesterol to the
pure lipid system. The second term corresponds to the ``chain
rigidifying'' effect reflecting the fact that a larger amount of
cholesterol favors ordered tail states and hence larger $\psi$.
Meanwhile, this coupling induces lipid-cholesterol phase
separation since it is proportional to $c^2$ with a negative
coefficient. In other words, the effective lipid-cholesterol
interaction depends on the conformational states of the lipid
chains.

Adding eqs.~(\ref{eq:flory}-\ref{eq:coupling}), we first minimize
$f^{\rm \ell c}$ with respect to $\psi$ and then construct the phase
diagrams as shown in fig.~\ref{lipidchol}.  In
fig.~\ref{lipidchol}(a), there are three different coexisting regions.
Due to the $\Gamma_1$ coupling term in eq.~(\ref{eq:coupling}), the
region of the disordered (D) phase, characterized by a non-zero small
value of $\psi$, widens at expense of ordered (O1) phase,
characterized by a large value of $\psi$, upon the addition of a small
amount of cholesterol ($c < 0.1$).  Note that D- and O1-phases
respectively correspond to \textit{L}$_d$ and gel phases in the
experiments.  For larger $c$, however, the $\Gamma_2$ coupling term
overcomes the first term, and the region of the D-phase narrows in
favor of O2-phase ($c > 0.1$), which corresponds to the \textit{L$_o$}
phase in which the order parameter $\psi$ takes the value much larger
than that in the D-phase.  The obtained phase diagram for
lipid-cholesterol systems agrees semiquantitatively with the
experimental one~\cite{VD,ST}.  The D- and O2-phases have the same
symmetry and are continuously connected, with a critical point at
$c_{\rm c} = 0.492 <0.5$. The appearance of the critical point can be
understood by noticing that the coupling terms in
eq.~(\ref{eq:coupling}) are linear in $\psi$, and act as an external
field coupled to $\psi$. In general, the first-order transition
becomes continuous when the applied external field is strong enough
\cite{GP}.

For larger $\Gamma_1$ and smaller $\Gamma_2$ we obtain the phase
diagram shown in fig.~\ref{lipidchol}(b). In this case, the
effective external field is too weak to eliminate the first-order
transition, and the critical point does not exist. This
phase diagram resembles that of binary lipid-lanosterol
mixtures~\cite{NTMIBZM}. Compared to cholesterol, lanosterol has
three additional methyl groups and has a structurally rougher
hydrophobic part, leading to a weaker enhancement of lipid
tail stretching (smaller $\Gamma_2$). Moreover, an extra double
bond in the lanosterol hydrocarbon tail can be expected to inhibit
crystallinity (larger $\Gamma_1$).
In fact, lanosterol is considered to be a precursor to cholesterol
in the evolutionary pathway~\cite{NTMIBZM}. Although cholesterol
can stabilize the region of coexistence between the D- and 
O2-phases (\textit{i.e.}, \textit{L$_d$}+\textit{L$_o$})
as shown in fig.~\ref{lipidchol}(a), such a
coexistence has not been identified for the lipid-lanosterol
mixtures~\cite{NTMIBZM}. The fact that the coexistence region ends
at $c=0.5$ is due only to the artificial truncation of the model
at $c=0.5$. If free excess lanosterol were allowed, the
phase boundaries would continue until the membrane lost its
integrity. However, such a break up of the membrane is not
considered here.

\section{Discussion}

Within a few simplifying assumptions we have obtained phase diagrams
that successfully mimic the experimental ones. A few points merit
further discussion. We used only one order parameter $\psi$ (related
to local membrane thickness) to describe the membrane internal degrees
of freedom of chain ordering and crystallinity. This assumption is
well motivated for pure lipid membrane where both ordering and
crystallization occur simultaneously. Although addition of cholesterol
may separate the chain ordering and freezing transitions and stabilize
the \textit{L}$_o$ phase, the thickness $\psi$ is sufficient here to
distinguish between the gel (O1), \textit{L}$_o$ (O2), and
\textit{L}$_d$ (D) phases. Note that the symmetry of the D- and
O2-phases is the same (hence the critical point), and differs from the
solid-like and local crystalline order of the O1-phase. As an
extension to the present model, it may be of interest to use two
distinct order parameters.

Finally, we discuss the qualitative behavior of the full ternary
phase diagram for saturated (S) and unsaturated (U) lipids in
the presence of cholesterol (C).  The phase diagrams in
figs.~\ref{lipidlipid}(a) and \ref{lipidchol}(a) correspond to
the two side views of the
schematic phase prism in fig.~\ref{schematic}(a), namely the
US-$T$ and the SC-$T$ planes, respectively.  A similar phase
diagram to fig.~\ref{lipidchol}(a) is expected for the UC-$T$
plane.  Consider a cut of the phase prism at a fixed temperature
as shown in fig.~\ref{schematic}(a).  Both the D+O coexistence
region on the US-$T$ plane and the O1+O2 coexistence region on the
SC-$T$ plane extend onto this triangle plane, and may form a
three-phase coexisting region (D+O1+O2, \textit{i.e.},
\textit{L$_d$}+gel+\textit{L$_o$}) when they meet each other.  If
the phase separation does not occur between U and C for this
temperature, a critical point between D and O2 is expected to
appear.  As the temperature is increased, the line of critical
points may end when the three-phase region disappears. These
arguments are consistent with recent experimental studies of the
phase morphology inside the phase prism~\cite{SGL,FB,VK,BHW}.
Further calculations of the full ternary phase diagrams are in
progress.

In summary, we have analyzed theoretically  saturated and
unsaturated lipid-lipid mixtures, and lipid-cholesterol mixtures.
The coupling between composition and the internal degree of
freedom (structural chain melting transition) allows us to obtain
phase diagrams that agree well with experiment, and to explain
some of the major features leading to creation of rafts.

\acknowledgments

We thank T. Kato for useful discussions, and the Isaac Newton
Institute, University of Cambridge, for its hospitality. 
This work is supported by the Japan Society for the Promotion
of Science, the Royal Society, the Ministry of Education, 
Culture, Sports, Science and Technology, Japan
under grant No.\ 15540395, the Israel Science Foundation (ISF) 
under grant No.\ 210/02, and the US-Israel Binational Foundation 
(BSF) under grant No.\ 287/02.


\end{document}